\documentclass[12pt]{article}
\usepackage{epsfig}
\usepackage{amsfonts}
\begin{document}
\begin{center}

{\Large Superstatistics: Theoretical concepts and 
physical applications}
%\chapter{Superstatistics: Theoretical concepts and physical applications}
%\label{ch1}
%\chapterauthor[Christian Beck]{Christian Beck}
% {First Author, Second Author\protect\footnote{Corresponding
%authors.}, and Third Author\protect\footnotemark[1]}

\vspace{0.5cm}

Christian Beck

\vspace{0.5cm}

School of Mathematical Sciences, Queen Mary, University of London,
London E1 4NS, UK
\end{center}

\vspace{0.5cm}

\abstract{A review of the superstatistics concept is provided, including
various recent applications to complex systems.}

\section{Introduction}
\label{ch1:sec1} Complex systems often exhibit a dynamics that
can be regarded as 
a superposition of  several dynamics on different time scales. As
a very simple example consider e.g.\ a Brownian particle moving
through a changing environment. Assume that the environment
exhibits temperature fluctuations on a large scale. Then there is
a relatively fast dynamics given by the velocity of the Brownian
particle and a slow one given by the temperature changes of the
environment, which is spatio-temporally inhomogeneous. The two
effects produce a superposition of two statistics, or in a short,
a `superstatistics' \cite{beck-cohen,BCS,supergen,touchette-beck,
souza,chavanis,plastino,abc}. The concept of a superstatistics was
introduced in \cite{beck-cohen}, in the mean time many
applications for a variety of complex systems have been pointed
out
\cite{beck03,reynolds,prl,beck-physica-d,maya,daniels,cosmic,rapisarda,bouchard,ausloos,
eco,abul,abe-thurner,hydro,rajagopal,briggs}. The stationary
probability distributions of superstatistical systems typically
exhibit non-Gaussian behaviour with fat tails, which can decay
e.g. with a power law, or as a stretched exponential, or in an
even more complicated way.

Essential for the superstatistical approach is the existence of an
intensive parameter $\beta$ which fluctuates on a large
spatio-temporal scale. For the above simple example of a
superstatistical Brownian particle, $\beta$ is the fluctuating
inverse temperature of the environment, but in general $\beta$
can also be an effective friction constant, a changing mass
parameter, a changing amplitude of Gaussian white noise, the
fluctuating energy dissipation in turbulent flows, a fluctuating
volatility in finance, an environmental parameter for biological
systems, or simply a local variance parameter extracted from a
signal. Some superstatistical models exhibit anomalous transport,
others don't.
Most superstatistical models
are somewhat `less anomalous' than Levy-type models, in the sense
that usually more of the higher moments exist as compared to Levy
processes. The tails of the distributions exhibit `fat'
tails, but usually these are less pronounced than for a Levy distribution.

%If $\beta$ is distributed
%according to a particular probability distribution, the
%$\chi^2$-distribution, then the corresponding marginal
%distributions, obtained by integrating over all $\beta$, are
%given by the generalized canonical distributions of nonextensive
%statistical mechanics
%\cite{tsa1,tsa2,tsa3,abe,souza,souza2,chavanis}. For other
%distributions of the intensive parameter $\beta$, one ends up
%with more general statistics, which contain Tsallis statistics as
%a special case.
The superstatistics concept is very general and has been applied to
a variety of complex systems. Recent successful applications
include hydrodynamic turbulence
\cite{beck03,reynolds,BCS,prl,beck-physica-d}, pattern forming systems
\cite{daniels}, cosmic rays\cite{cosmic}, solar
flares\cite{maya}, mathematical finance
\cite{bouchard,ausloos,eco}, random matrix theory \cite{abul},
networks \cite{abe-thurner}, quantum systems at low temperatures
\cite{rajagopal}, wind velocity fluctuations
\cite{rapisarda,rap2}, hydro-climatic fluctuations \cite{hydro}
and delay statistics in traffic models \cite{briggs}. The aim in
the following is to explain the basic concepts and applications
in an easy-going way.

\section{The basic idea}

Consider a complex system in a stationary nonequilibrium state
that is driven by some external forces. Usually we think here of
a physical system (e.g.\ a turbulent flow) but we may easily
apply similar techniques to economic, biological, social systems,
where the meaning of the mathematical variables will be
different, though the mathematical structure is similar.
Generally, a complex system will be inhomogeneous in space and in
time. Effectively, it may consist of many spatial cells (or, the
measured time series may consist of many time slices) where there
are different values of some relevant system parameter $\beta$.
The cell size is effectively determined by the correlation length
of the continuously varying $\beta$-field. Superstatistical
systems are characterized by a simplifying effect, namely the
fact that the relaxation time is short so that each cell can be
assumed to be in local equilibrium (in a certain approximation at
least). Sometimes this property will be satisfied for a given
complex system, sometimes not.

In the long-term run, the stationary distributions of the
superstatistical inhomogeneous system arise as a superposition of
Boltzmann factors $e^{-\beta E}$ (or analogues of Boltzmann
factors describing the local behaviour of the system under
consideration) weighted with the probability density $f(\beta)$
to observe some value $\beta$ in a randomly chosen cell:
\begin{equation}
p(E)=\int_0^\infty f(\beta)  \frac{1}{Z(\beta)} \rho(E) e^{-\beta
E}d\beta  \label{ppp}
\end{equation}
Here $E$ is an effective energy for each cell, $Z(\beta)$ is the
normalization constant of $\rho(E) e^{-\beta E}$ for a given
$\beta$, and $\rho (E)$ is the density of states.

A typical example is a Brownian particle of mass $m$ moving
through a changing environment in $d$ dimensions. Such a Langevin
model has applications in many different areas of science. In the
simplest case we may write down a linear local Langevin equation
for the velocity $\vec{v}$
\begin{equation}
\dot{\vec{v}}=-\gamma \vec{v} + \sigma \vec{L}(t)
\end{equation}
($\vec{L}(t)$: $d$-dimensional Gaussian white noise) which
becomes superstatistical due to the fact that for a fluctuating
environment the parameter $\beta :=\frac{2}{m}
\frac{\gamma}{\sigma^2}$ becomes a random varaible as well: It
varies from cell to cell on a rather large spatio-temporal scale
$T$ (this time scale $T$ should not be confused with the
temperature, which is denoted as $\beta^{-1}$). Of course, for
this example $E=\frac{1}{2}mv^2$, and while the local stationary
distribution in each cell is Gaussian
\begin{equation}
p(\vec{v}|\beta)=\left( \frac{\beta}{2\pi}\right)^{d/2}
e^{-\frac{1}{2}\beta mv^2},
\end{equation}
the marginal distribution describing the long-time behaviour of
the particle
\begin{equation}
p(\vec{v})=\int_0^\infty f(\beta)p(\vec{v}|\beta)d\beta
\label{margi}
\end{equation}
exhibits non-trivial behaviour. The large-$|v|$ tails of this
distribution depend on the behaviour of $f(\beta )$ for $\beta
\to 0$ \cite{touchette-beck}. As a result of the integration over
$\beta$, the probability distribution $p(\vec{v})$ will typically
have fat tails. These can be e.g.\ power law tails, stretched
exponentials, or whatever. 

One of the most important example for
practical application is a $\chi^2$-distribution for $f(\beta)$
(see next section for concrete formulas) and a Gaussian
distribution for $p(\vec{v}|\beta)$. In this case one obtains
from eq.~(\ref{margi})
\begin{equation}
p(\vec{v})\sim \frac{1}{(1+(q-1)\frac{1}{2}bv^2)^{\frac{1}{q-1}}},
\label{qgau}
\end{equation}
where $q$ and $b$ are suitable parameters, and $v=|\vec{v}|$. The
function on the right-hand side of eq.~(\ref{qgau}) is called a
$q$-Gaussian \cite{abe} and denoted by $e_q^{-\frac{1}{2}bv^2}$. Note that
$q$-Gaussians decay asymptotically with a power law for $v \to
\infty$. They reduce to ordinary Gaussians for $q\to 1$. But
power laws are not the only possibility one can get
for superstatistical systems. Many examples
will be discussed in the following sections.

A generalized thermodynamics for superstatistical systems has
been recently developed in \cite{abc} (see also \cite{souza} for
early attempts). Here one starts from 
a Boltzmann-Gibbs-Shannon entropy function that also includes contributions from the
fluctuations in $\beta$. Given that entropy, one can do formally
thermodynamics. Ordinary thermodynamics (no fluctuations in
$\beta$) is contained as a special case in this more general
formalism.

\section{Typical distributions $f(\beta)$}
The distribution $f(\beta)$ is determined by the spatio-temporal
dynamics of the driven nonequilibrium system under consideration. By construction,
$\beta$ is positive, so $f(\beta)$ cannot be a Gaussian. Let us
here consider important examples of what to expect in typical
experimental situations for driven nonequilibrium systems.

a) There may be many (nearly) independent microscopic random
variables variables $\xi_j$, $j=1,\ldots , J$ contributing to
$\beta$ in an additive way. For large $J$ their rescaled sum
$\frac{1}{\sqrt{J}}\sum_{j=1}^J\xi_j$ will approach a Gaussian
random variable $X_1$ due to the Central Limit Theorem (CLT). In
total, there can be many subsystems consisting of such
microscopic random variables, leading to $n$ Gaussian random
variables $X_1,\ldots ,X_n$ due to various degrees of freedom in
the system. As mentioned before, $\beta$ needs to be positive and
a positive $\beta$ is obtained by squaring these Gaussian random
variables. The resulting $\beta=\sum_{i=1}^NX_i^2$ is
$\chi^2$-distributed with degree $n$, i.e.
\begin{equation}
f(\beta )=\frac 1{\Gamma (\frac n2)}\left( \frac n{2\beta
_0}\right) ^{n/2}\beta ^{n/2-1}e^{-\frac{n\beta }{2\beta _0}}.
\label{chi22}
\end{equation}
The marginal
distributions obtained by integrating over all $\beta$ exhibit
power-law tails for large enerfies $E$. They are 
$q$-exponentials, $p(E)\sim e_q^{-bE}=(1+(q-1)bE)^{-1/(q-1)}$, where $q$ and $b$ can be
related to $n$ and $\beta_0$ \cite{wilk,prlold}. Note that this
statistics arises as a universal limit dynamics, i.e.\ the
details of the microscopic random variables $\xi_j$ (e.g.\ their
probability densities) are irrelevant.

b) The same consideration as above may apply to the `temperature'
$\beta^{-1}$ rather than $\beta$ itself. $\beta^{-1}$ may the sum
of several squared Gaussian random variables arising out of many
microscopic degrees of freedom $\xi_j$. The resulting $f(\beta)$
is the inverse $\chi^2$-distribution given by
\begin{equation}
f(\beta )=\frac{\beta _0}{\Gamma (\frac n2)}\left( \frac{n\beta
_0}2\right) ^{n/2}\beta ^{-n/2-2}e^{-\frac{n\beta _0}{2\beta }}.
\end{equation}
It generates distributions that have exponential decays in
$\sqrt{E}$ \cite{touchette,touchette-beck,sattin}. Again this
superstatistics is universal: details of the $\xi_j$ are
irrelevant.

c) Instead of $\beta$ being a sum of many contributions, for
other systems (in particular turbulent ones) the random variable
$\beta$ may be generated by multiplicative random processes. We
may have a local cascade random variable $X_1= \prod_{j=1}^{J}
\xi_j$, where $J$ is the number of cascade steps and the $\xi_i$
are positive microscopic random variables. Due to the CLT, 
$\log X_1= \sum_{j=1}^J \log \xi_j$ becomes Gaussian for large $J$ if
it is properly rescaled. Hence $X_1$ is lognormally distributed.
In general there may be $n$ such product contributions to the
superstatistical varaiable $\beta$, $\beta = \prod_{i=1}^n X_i$.
Then $\log \beta = \sum_{i=1}^n \log X_i$ is a sum of Gaussian
random variables, hence it is Gaussian as well. Thus $\beta$ is
lognormally distributed, i.e.
\begin{equation}
f(\beta )=\frac a\beta \exp \left[ -c(\ln \beta -b)^2\right] ,
\end{equation}
where $a,b,c$ are suitable constants.
The result is independent of the details of the microscopic
cascade random variables $\xi_j$, hence there is universality
again. This type of lognormal superstatistics is particularly relevant
for turbulent flows \cite{beck-physica-d, reynolds, beck03,
castaing, prl}.

\section{Asymptotic behaviour for large energies}

Superstatistical probability densities, as given by
eq.(\ref{ppp}) or (\ref{margi}), typically exhibit `fat tails'
for large $E$, but what is the precise functional form of this
large energy behaviour? The answer depends on the distribution
$f(\beta)$ and can be obtained from a variational principle.
Details are described in \cite{touchette-beck}, here we just
summarize some results. For simplicity, let us put $\rho (E)=1$ in
eq.~(\ref{ppp}). We may define a new probability density
$\tilde{f}$ by
\begin{equation}
\tilde{f} (\beta):=c\frac{f(\beta)}{Z(\beta)},
\end{equation}
where $c$ is a suitable normalization constant. The new density
$\tilde{f}$ absorbes the $\beta$-dependence of the local partition
function $Z(\beta)$. With this notation, $p(E)$ can now be regarded as
the Laplace transform of $\tilde{f}$. Renaming $\tilde{f}\to f$ we
obtain
\begin{eqnarray}
p(E) &\sim&\int_0^\infty f(\beta )e^{-\beta E}d\beta  \nonumber \\
&=&\int_0^\infty e^{-\beta E+\ln f(\beta )}d\beta  \nonumber \\
&\sim &e^{\sup_\beta \{-\beta E+\ln f(\beta )\}}  \nonumber \\
&=&e^{-\beta _EE+\ln f(\beta _E)}  \nonumber \\
&=&f(\beta _E)e^{-\beta _EE}. \label{5}
\end{eqnarray}
Here we used the saddle point approximation. $\beta_E$ is the
value of $\beta$ where the function $-\beta E+\ln f(\beta )$ has
a maximum. The expression \begin{equation} \sup_\beta \{-\beta
E+\ln f(\beta )\}
\end{equation}
corresponds to a Legendre transform of $\ln f(\beta )$.
%The
%result of this transform is a function of $E$ which can be
%thought of as representing a kind of entropy function if we
%consider the function $\ln f(\beta )$ to represent a free energy
%function. This entropy function, however, is different from other
%entropy functions used e.g.\ in nonextensive statistical
%mechanics. It describes properties related to the fluctuations of
%inverse temperature.

For the case where $f(\beta )$ is smooth and has only a single
maximum we can obtain the supremum by differentiating, i.e.\
\begin{equation}
\sup_\beta \{-\beta E+\ln f(\beta )\}=-\beta _EE+\ln f(\beta _E)
\end{equation}
where $\beta _E$ satisfies the differential equation
\begin{equation}
0=-E+(\ln f(\beta ))^{\prime }=-E+\frac{f^{\prime }(\beta
)}{f(\beta )}. \label{lf1}
\end{equation}
By taking into account the next-order contributions around the
maximum, eq.~(\ref{5}) can be improved to
\begin{equation}
p(E)\sim \frac{f(\beta _E)e^{-\beta _EE}}{\sqrt{-(\ln f(\beta
_E))^{\prime \prime }}}.
\end{equation}

Let us consider a few examples. Consider an $f(\beta)$ which
for small $\beta$ is of the
power-law form $f(\beta )\sim \beta ^\gamma $, $\gamma >0$.
An example is the $\chi ^2$-distribution of $n$
degrees of freedom, which was mentioned previously:
\begin{equation}
f(\beta )=\frac 1{\Gamma (\frac n2)}\left( \frac n{2\beta
_0}\right) ^{n/2}\beta ^{n/2-1}e^{-\frac{n\beta }{2\beta _0}},
\label{chi2}
\end{equation}
($\beta _0\geq 0$, $n>1$). This behaves for $\beta \to 0$ as
\begin{equation}
f(\beta )\sim \beta ^{n/2-1},
\end{equation}
i.e.\
\begin{equation}
\gamma = \frac{n}{2}-1.
\end{equation}
Other examples exhibiting this power-law form are the so-called
$F$-distributions\cite{beck-cohen,sattin}. With the above
formalism one obtains from eq.~(\ref{lf1})
\begin{equation}
\beta _E =\frac \gamma E
\end{equation}
and
\begin{equation}  p(E) \sim E^{-\gamma -1}.
\end{equation}
These types of $f(\beta)$ form the basis for power-law
generalized Boltzmann factors ($q$-exponentials) used in
generalized versions of statistical mechanics, so-called
non-extensive statistical mechanics \cite{tsa1,tsa2,tsa3,abe}.
These depend on an entropic index $q$ coming from a more general
entropy functional, and the relation between $\gamma$ and $q$ is
\begin{equation}
\gamma +1 = \frac{1}{q-1}. \label{here}
\end{equation}

Another example would be an $f(\beta)$ which for small $\beta$
behaves as $f(\beta )\sim e^{-c/\beta }$, $c>0$. In this case one
obtains
\begin{equation}
\beta_E=\sqrt{\frac{c}{E}}
\end{equation}
and \begin{equation}
 p(E) \sim
E^{-3/4}e^{-2\sqrt{cE}}.
\end{equation}
The above example can be generalized to stretched exponentials:
For $f(\beta)$ of the form $f(\beta )\sim e^{-c\beta ^\delta }$
one obtains after a short calculation
\begin{equation}
\ \beta _E =\left( \frac E{c|\delta |}\right) ^{1/(\delta -1)}
\end{equation}
and
\begin{equation}
p(E) \sim E^{(2-\delta )/(2\delta -2)}e^{aE^{\delta /(\delta
-1)}},
\end{equation}
where $a$ is some factor depending on $\delta$ and $c$.
In this case the superstatistical complex system exhibits stretched exponential tails.

\section{Anomalous diffusion in superstatistical systems}

We now illustrate that superstatistical systems can exhibit
normal as well as anomalous transport. This depends on the
dynamical properties of the model considered.

For simplicity, we restrict ourselves to a 1-dimensional model.
Let us again consider locally a 1-dimensional Brownian particle of mass
$m$ and a Langevin equation of the form
\begin{equation}
\dot{v}=-\gamma v + \sigma L(t),
\end{equation}
where $v$ denotes the velocity of the particle, and $L(t)$ is
normalized Gaussian white noise with the following expectations:
\begin{eqnarray}
\langle L(t) \rangle &=&0 \\
\langle L(t)L(t') \rangle &=&\delta (t-t').
\end{eqnarray}
We assume that the parameters $\sigma$ and $\gamma$ are constant
for a sufficiently long time scale $T$, and then change to new
values, either by an explicit time dependence, or by a change of
the environment through which the Brownian particle moves. Formal
identification with local equilibrium states in the spatial cells
where $\beta$ is approximately constant
(ordinary statistical mechanics at temperature $\beta^{-1}$)
yields during the time scale $T$ the relation\cite{vKa}
\begin{equation}
\langle v^2 \rangle =\frac{\sigma^2}{2\gamma}=\frac{1}{\beta m}
\label{einstein}
\end{equation}
or
\begin{equation}
\beta = \frac{2}{m} \frac{\gamma}{\sigma^2}.  \label{beta}
\end{equation}
Again, we emphasize that after the time scale $T$, $\gamma$ and
$\sigma$ will take on new values in a stochastic
way. During a time interval of the
order of $T$, the probability density $P(v,t)$ obeys the
Fokker-Planck equation
\begin{equation}
\frac{\partial P}{\partial t}=\gamma \frac{\partial (vP)}{\partial
v}+ \frac{1}{2} \sigma^2 \frac{\partial^2 P}{\partial v^2}
\end{equation}
with the local stationary solution
\begin{equation}
P(v|\beta)=\sqrt{\frac{m\beta}{2\pi}} \exp \left\{-\frac{1}{2}
\beta mv^2 \right\} \label{28}.
\end{equation}
In the adiabatic approximation, valid for large $T$, one asumes
that the local equilibrium state is reached very fast so that
relaxation processes can be neglected. Within a cell in local
equilibrium the correlation function is given by \cite{vKa}
\begin{equation}
C(t-t'|\beta)=\langle v(t) v(t')\rangle =
\frac{1}{m\beta}e^{-\gamma |t-t'|}.
\end{equation}
%Clearly, for $t=t'$ and setting $m=1$ we have
%\begin{equation}
%\beta= \frac{1}{\langle v^2 \rangle_T},
%\end{equation}
%where the index $T$ indicates that the average is performed over
%the time interval $T$ only.

It is now interesting to see that the long-term invariant
distribution $P(v)$, given by
\begin{equation}
P(v)=\int_0^\infty f(\beta) P(v|\beta) d\beta \label{31}
\end{equation}
depends only on the probability distribution of
$\beta=\frac{2}{m} \frac{\gamma}{\sigma^2}$ and not on that of the
single quantities $\gamma$ and $\sigma^2$. This means, one can
obtain the same stationary distribution (\ref{31}) from different dynamical
models based on a Langevin equation with fluctuating parameters.
Either $\gamma$ may fluctuate, and $\sigma^2$ is constant, or the
other way round. On the other hand, the superstatistical
correlation function
\begin{equation}
C(t-t')=\int_0^\infty f(\beta) C(t-t'|\beta) d\beta =\frac{1}{m}
\int_0^\infty f(\beta) \beta^{-1}e^{-\gamma |t-t'|}d\beta
\label{32}
\end{equation}
can distinguish between these two cases. The study of correlation
functions thus yields more information for any superstatistical
model.

Let us illustrate this with a simple example. Assume that $\sigma$
fluctuates and $\gamma$ is constant and that $\beta=\frac{2}{m}
\frac{\gamma}{\sigma^2}$ is $\chi^2$-distributed. Since $\gamma$
is constant, we can move the exponential $e^{-\gamma |t-t'|}$ out
of the integral in eq.~(\ref{32}), meaning that the
superstatistical correlation function still decays in an
exponential way:
\begin{equation}
C(t-t') \sim e^{-\gamma |t-t'|}. \label{expo}
\end{equation}
On the other hand, if $\sigma$ is constant and $\gamma$ fluctuates
and $\beta$ is still $\chi^2$-distributed with degree $n$, we get
a completely different answer. In this case, in the adiabatic
approximation, the integration over $\beta$ yields a power-law
decay of $C(t-t')$:
\begin{equation}
C(t-t') \sim |t-t'|^{-\eta}, \label{power}
\end{equation}
where
\begin{equation}
\eta =\frac{n}{2}-1 \label{eta}
\end{equation}
Note that this decay rate is different from the asymptotic power
law decay rate of the invariant density $P(v)$, which, using
(\ref{28}) and (\ref{31}), is given by $P(v)\sim v^{-2/(q-1)}$,
with
\begin{equation}
\frac{1}{q-1} =\frac{n}{2}+\frac{1}{2}. \label{qqq}
\end{equation}

Now let us proceed to the position
\begin{equation}
x(t)=\int_0^t v(t')dt'
\end{equation}
of the test particle. One has
\begin{equation}
\langle x^2(t) \rangle = \int_0^t \int_0^t \langle
v(t')v(t'')\rangle dt'dt''.
\end{equation}
Asymptotic power-law velocity correlations with an exponent
$\eta <1 $ are expected to imply asymptotically anomalous
diffusion of the form
\begin{equation}
\langle x^2 (t) \rangle  \sim t^\alpha \label{alpha}
\end{equation}
with
\begin{equation}
\alpha =2-\eta .
\end{equation}
This relation simply results from the two time integrations.

It is interesting to compare our superstatistical model with other dynamical models
generating anomalous diffusion. Plastino and Plastino\cite{plasti}
and Tsallis and Bukmann\cite{bukmann} study a generalized
Fokker-Planck equation of the form
\begin{equation}
\frac{\partial P(x,t)}{\partial t}=- \frac{\partial}{\partial x}
(F(x)P(x,t))+ D \frac{\partial^2}{\partial x^2}P(x,t)^\nu
\label{buk}
\end{equation}
with a linear force $F(x)=k_1-k_2x$ and $\nu\not= 1$. Basically
this model means that the diffusion constant becomes dependent on
the probability density $P$. The probability densities generated by
eq.~(\ref{buk}) are $q$-Gaussians with the exponent
\begin{equation}
q=2-\nu.
\end{equation}
The model generates anomalous diffusion with $\alpha = 2/(3-q)$.
Assuming the validity of $\alpha = 2-\hat{\eta}$, i.e. the
generation of anomalous diffusion by slowly decaying velocity
correlations with exponent $\hat{\eta}$, one obtains
\begin{equation}
\hat{\eta} =\frac{4-2q}{3-q}.
\end{equation}
On the other hand, for the $\chi^2$-superstatistical Langevin
model one obtains by combining eq.~(\ref{eta}) and (\ref{qqq}) the
different relation
\begin{equation}
\eta =\frac{5-3q}{2q-2}.
\end{equation}
Interesting enough, there is a distinguished $q$-value where both
models yield the same answer:
\begin{equation}
q=1.453 \Rightarrow \hat{\eta} =\eta =0.707
\end{equation}
These values of $q$ and $\eta$ correspond to realistic,
experimentally observed numbers, for example in defect
turbulence (see section 9).

\section{From time series to superstatistics}

We now want to be more practically orientated and apply
superstatistical techniques to some complex systems (of whatever kind) where we do
not know the equations of motion, and neither the distribution
$f(\beta)$, but do have some information in form of a measured
time series. Suppose an experimentally measured scalar time series
$u(t)$ is given. Our goal is to test the hypothesis that it is
due to a superstatistics and if yes, to extract $f(\beta)$. First
we have to determine the superstatistical time scale $T$. For
this we divide the time series into $N$ equal time intervals of
size $\Delta t$. The total length of the signal is $t_{max}=N\Delta t$. We
then define a function $\kappa(\Delta t)$ by
\begin{equation}
\kappa (\Delta t) = \frac{1}{t_{max}-\Delta t}  \int_0^{t_{max}-\Delta t} dt_0 \frac{\langle
(u-\bar{u})^4 \rangle_{t_0,\Delta t}}{\langle (u-\bar{u})^2
\rangle^2_{t_0,\Delta t}}
\end{equation}
Here $\langle \cdots \rangle_{t_0,\Delta t}=\frac{1}{\Delta t}
\int_{t_0}^{t_0+\Delta t}\cdots dt$ denotes an average over an
interval of length $\Delta t$ starting at $t_0$. The integration result
fluctuates for each value of $t_0$ and is averaged by the
integral over $t_0$. $\bar{u}$ denotes the average of $u$.
Now assume the simplest case, that our
complex system dynamics basically arises out of a superposition of
local Gaussians on some unknown time scale $T$. How
can we extract $T$? We should be looking for the special value
$\Delta t =T$ where
\begin{equation}
\kappa (T)= 3.
\end{equation}
Clearly this condition defining the superstatistical time scale
$T$ simply reflects the fact that we are looking for locally
Gaussian behaviour in the time series, which implies a local
flatness of 3. If $\Delta t$ is so small that only one constant value
of $u$ is observed in this interval, then of course
$\kappa(\Delta t)=1$. On the other hand, if $\Delta t$ is so large that
it includes the entire time series, then we obtain the flatness
of the distribution of the entire signal, which is larger than 3,
since superstatistical distributions are generically fat-tailed.
Inbetween, there should be a distinguished time scale where
$\kappa=3$.

Fig.~1 shows the function $\kappa (\Delta t)$ for an example of a
time series that has been studied in \cite{BCS}, the longitudinal
velocity difference $u(t)=v(t+\delta) -v(t)$ in a turbulent
Taylor-Couette flow on a scale $\delta$.
\begin{figure}
\epsfig{file=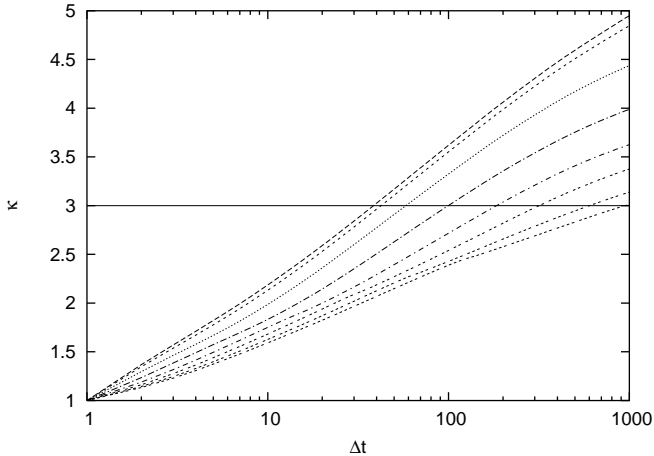} 
\caption{Determination of the
superstatistical time scale $T$ from the intersection with the line $\kappa =3$
for turbulent Taylor-Couette flow, $\delta =2^j,j=0,1,2,\ldots, 7$ (from top to bottom).}
\end{figure}
For each scale $\delta$ the relevant superstatistical time scale
$T$ leading to locally Gaussian behaviour can be extracted as the
intersection with the line $\kappa =3$. These time scales $T$ are
to be compared with the relaxation times $\gamma^{-1}$ of the
dynamics, which can be estimated from the short-time exponential
decay of the correlation function $C(t-t'))=\langle
u(t)u(t')\rangle$. One obtains the result that the ratio
$T/\gamma^{-1}$ is pretty large and increases with Reynolds
number \cite{BCS}. This time scale separation is indeed the deeper reason
why superstatistical models of turbulence work quite well
\cite{beck-physica-d,BCS,prl}.

Next, given a general signal $u(t)$ we are interested in the
analysis of the slowly varying stochastic process $\beta (t)$.
Since the variance of local Gaussians $\sim e^{-\frac{1}{2}\beta
u^2}$ is given by $\beta^{-1}$, we can determine the process
$\beta (t)$ from the time series as
\begin{equation}
\beta (t_0) = \frac{1}{\langle u^2 \rangle_{t_0,T}-\langle u
\rangle^2_{t_0,T}}
\end{equation}
We can then easily make a histogram of $\beta (t_0)$ for all
values of $t_0$, thus obtaining the probability density
$f(\beta)$.

Fig.~2 shows this probability density for our example of 
turbulent time series.
\begin{figure}
\epsfig{file=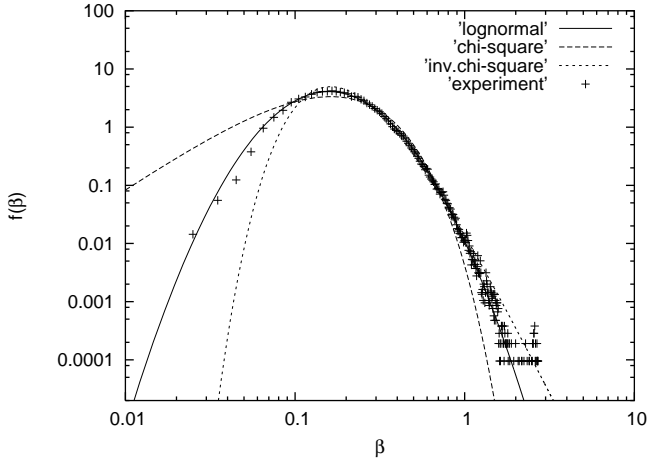} \caption{Probability distribution
$f(\beta)$ as extracted from the measured turbulent time series
in \cite{BCS}.}
\end{figure}
Motivated by our consideration in section 3, the data are
compared with a $\chi^2$-distribution, inverse
$\chi^2$-distribution and lognormal distribution, all having the
same mean $\langle \beta \rangle$ and variance $\langle
\beta^2\rangle -\langle \beta \rangle^2$ as the experimental
data. Clearly the lognormal distribution yields the best
fit. Indeed, the cascade picture of energy dissipation in
turbulent flow strongly suggests that 
lognormal superstatistics should be relevant, with $\beta$
being related to a suitable power of the fluctuating energy dissipation rate
\cite{prlold,prl}.
However, again let us mention that other complex
systems can generate a completely different type of superstatistics.
Still the same methods apply.

For superstatistics to be a good approximation we need the
variable $\beta (t)$ to change very slowly as compared to $u(t)$. This
is indeed the case for our turbulence example, as can be seen in Fig.~3.
\begin{figure}
\epsfig{file=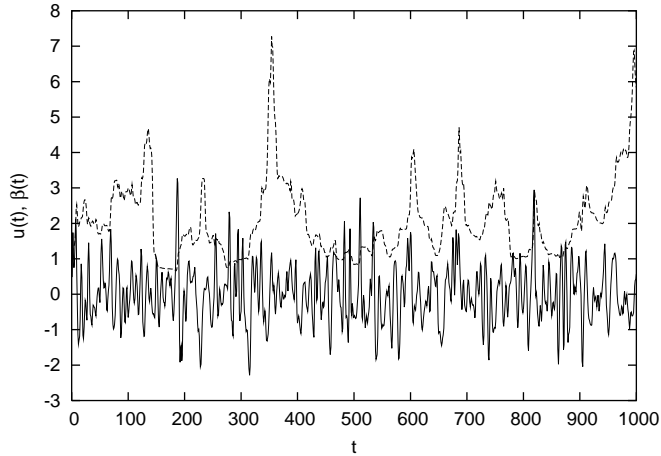} \caption{Typical evolution of
$u(t)$ (solid line) and $\beta (t)$ (dashed line) for the turbulent Taylor-Couette flow.}
\end{figure}

\section{Overview of applications}

The superstatistics concept has been applied to many different
complex systems. Let us first give a short overview of
applications, after that we will treat four examples in more
detail. Rizzo and Rapisarda\cite{rapisarda,rap2} study
experimental data of wind velocities at Florence airport and find
that superstatistics does a good job. Swinney et al.
\cite{BCS,jung-swinney} extract lognormal superstatistics for
turbulent flow between counterrotating disks.
%Jung and
%Swinney\cite{jungswinney} study velocity differences in a
%turbulent Taylor-Couette flow, which is well described by
%lognormal superstatistics. They also find a simple scaling
%relation between the superstatistical parameter $\beta$ and the
%fluctuating energy dissipation $\epsilon$.
Paczuski et al.\cite{maya} study data of solar flares on various
time scales and embedd this into a superstatistical model based on
$\chi^2$-superstatistics. Human behaviour when sending off print
jobs might also stand in connection to such a
superstatistics\cite{pac2}. Bodenschatz et al.\cite{boden,boden1,boden2}
and Pinton et al. \cite{pinton1,pinton2} have detailed
experimental data on the accelerations of single test particles in
a turbulent flow, which are well described by lognormal
superstatistics \cite{beck03,reynolds,prl}. The statistics of cosmic
rays is well described by $\chi^2$-superstatistics, with $n=3$
due to the three spatial dimensions\cite{cosmic}. In mathematical
finance superstatistical techniques are well known and come under
the heading `volatility fluctuations', see
e.g.\cite{bouchard,ausloos,eco}. Possible applications also
include granular media, which could be described by different
types of superstatistics, depending on the boundary
conditions\cite{vanzon}. The observed fat tails of solar wind
speed fluctuations\cite{burlaga} could also be related to a
superstatistical model. Hydroclimatic fluctuations have been
analysed using the superstatistics concept \cite{hydro}.
Briggs et al. \cite{briggs} apply a superstatistical model to
observed train delays on the British rail network. On the
theoretical side, Chavanis\cite{chavanis} points out analogies
between superstatistics and the theory of violent relaxation for
collisionless stellar systems. Abul-Magd \cite{abul} applies
superstatistics to random matrix theory. 
Luczka and
Zaborek\cite{luczka} have studied a simple model of
dichotomous fluctuations of $\beta$ on different time scales
where everything can be
calculated analytically. Mathai and Haubold \cite{hau} investigate
a link between superstatistics and fractional reaction equations.

\section{Lagrangian turbulence}

We now treat some examples of applications in more detail. We
start with the recent Lagrangian turbulence applications
\cite{prl}. Over the past few years there has been experimental progress
\cite{boden,boden1,boden2,pinton1,pinton2} in tracking single test
particles advected by a turbulent flow. This area of research is called `Lagrangian
turbulence'. To theoretically model Lagrangian turbulence, one may
first start from a Gaussian turbulence model, the Sawford model
\cite{sawford, pope}. This model considers the joint stochastic
process $(a(t),v(t),x(t))$ of an arbitrary component of
acceleration, velocity and position of a Lagrangian
test particle embedded in the turbulent flow, and assumes that
they obey the stochastic differential equation
\begin{eqnarray}
\dot{a}& =&-(T_L^{-1}+t_\eta^{-1})a-T_L^{-1}t_\eta^{-1} v\nonumber
\\ &\,&
+\sqrt{2\sigma_v^2(T_L^{-1}+t_\eta^{-1})T_L^{-1}t_\eta^{-1}}\;
L(t)
\\ \dot{v} &=&a \\ \dot{x} &=&v,
\end{eqnarray}

where

 \indent $L(t)$: Gaussian white noise

$T_L$ and $t_{\eta}$:  two time scales, with $T_L
>>t_\eta$,

$T_L=2\sigma_v^2/(C_0 \bar{\epsilon})$

$t_\eta = 2a_0\nu^{1/2}/(C_0\bar{\epsilon}^{1/2})$

$\bar{\epsilon}$: average energy dissipation

$C_0, a_0$: Lagrangian structure function constants

$\sigma_v^2$ variance of the velocity distribution

$ R_\lambda = \sqrt{15}\sigma_v^2/\sqrt{\nu \bar{\epsilon}} $
Taylor scale Reynolds number.

For our purposes it is sufficient to consider the limit $T_L \to
\infty$, which is a good approximation for large Reynolds
numbers. In that limit the Sawford model reduces to just a linear
Langevin equation for the acceleration

\begin{equation}
\dot{a}=-\gamma a +\sigma L(t) \label{la}
\end{equation}
with
\begin{eqnarray}
\gamma &=&\frac{C_0}{2a_0} \nu^{-1/2} \bar{\epsilon}^{1/2}\label{gamma} \\
\sigma &=& \frac{C_0^{3/2}}{2a_0} \nu^{-1/2} \bar{\epsilon}.
\label{sigma}
\end{eqnarray}
Note that this is a Langevin equation for the acceleration, so the meaning
of the variables is slightly different as compared to the case of an ordinary
Brownian particle, where the Langevin equation describes the velocity.
In practice, the acceleration is measured as a velocity difference on a very
small time scale.

Unfortunately, the Sawford model predicts Gaussian stationary
distributions for $a$ and $v$, and is thus at variance with the
recent measurements \cite{boden1,pinton1}, which provide evidence for 
distributions with fat tails. So how can we extend the Sawford
model to make it physically realistic?

As said before, the idea is to generalize the Sawford model with
constant parameters to a superstatistical Sawford model with
fluctuating ones. To construct a superstatistical extension of
Sawford model, one replaces the constant energy dissipation
$\bar{\epsilon}$ by a fluctuating one. It is assumed to be lognormally
distributed. Moreover, one extends the model to include all 3
components of the velocity and acceleration, as well as
contributions from a fluctuating enstrophy (rotational energy)
surrounding the test particle. From this new theory \cite{prl} excellent agreement
with the experimental data is obtained, see 
Fig.~4 for an example.
\begin{figure}
\epsfig{file=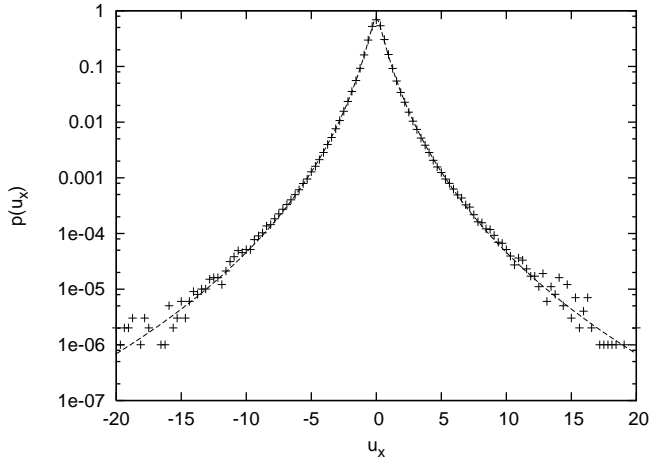} \caption{Predicted and measured
probability density of a component of the small-scale velocity difference
of a
Lagrangian test particle in a turbulent flow.
The dashed line is based on lognormal superstatistics \cite{prl}. The
experimental data are from \cite{pinton1}.}
\end{figure}
One obtains not only the correct 1-point probability
distributions, but also good agreement for the decay of
correlation functions, the observed statistical dependencies
between acceleration components, and scaling exponents.

\section{Defect turbulence}

Let us now consider another physically relevant example, so-called
`defect turbulence'. Defect turbulence shares with ordinary
turbulence only the name, otherwise it is very different. It is a
phenomenon related to convection and has nothing to do with fully
developed hydrodynamic turbulence. Consider a Raleigh-Benard
convection experiment: A liquid is heated from below and cooled
from above. For large enough temperature differences, interesting
convection patterns start to evolve. An inclined layer convection
experiment \cite{daniels1,daniels2,daniels} is a kind of Raleigh-Benard experiment where the
apparatus is tilted by an angle (say 30 degrees), moreover the
liquid is confined between two very narrow plates. For large
temperature differences, the convection rolls evolve
chaotically. Of particular interest are the defects in this
pattern, i.e.\ points where two convection rolls merge into one
(see Fig.~5).
\begin{figure}
\epsfig{file=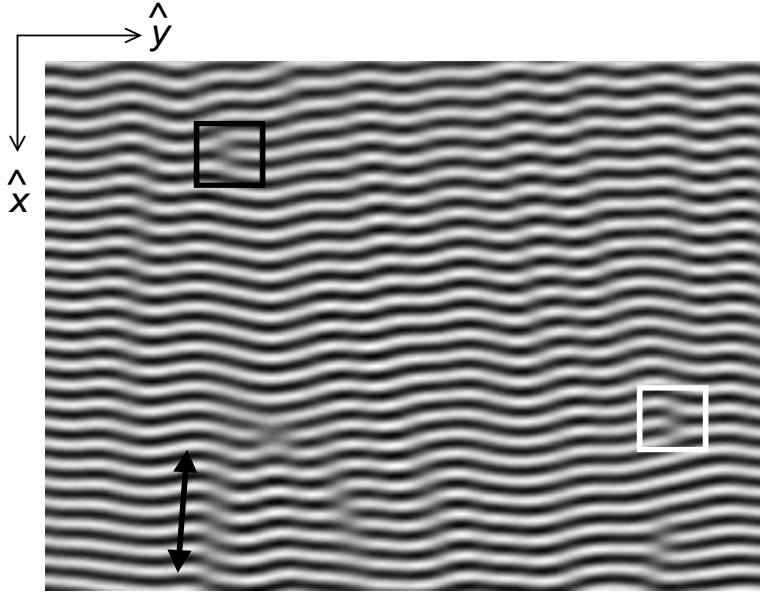, width=10cm, height=8cm} \caption{Convection rolls and
defects (black and white boxes) as observed in the experiment of Daniels et al.
\cite{daniels}.}
\end{figure}
These defects behave very much like particles. They have a
well-defined position and velocity, they are created and
annihilated in pairs, and one can even formally
attribute a `charge' to them: There are positive and negative
defects, as indicated by the black and white boxes in Fig.~5.

The probability density of defect velocities has been quite
precisely measured \cite{daniels}. As shown in Fig.~6, it quite precisely
coincides with a $q$-Gaussian with $q \approx 1.46$.
\begin{figure}
\epsfig{file=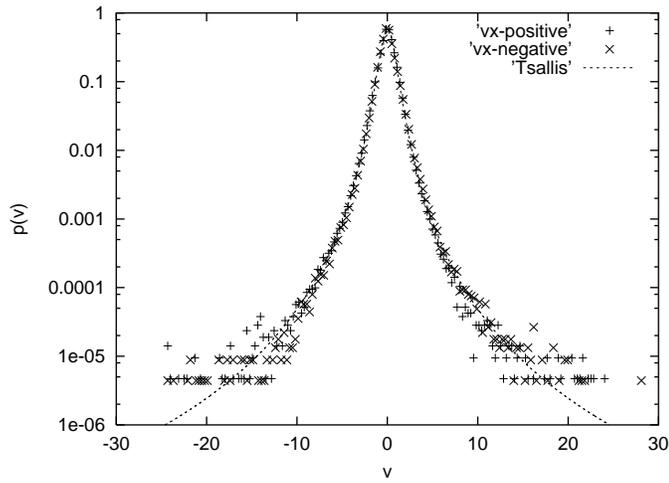} \caption{Measured probability
density of defect velocities and fit with a $q$-Gaussian with $q=1.46$.}
\end{figure}
The defects are also observed to exhibit anomalous diffusion. Their
position $X(t)$ roughly obeys an anomalous diffusion law of the type
\begin{equation}
\langle X^2(t) \rangle \sim t^\alpha,
\end{equation}
where $\alpha \approx 1.33$ (see Fig.~7).
\begin{figure}
\epsfig{file=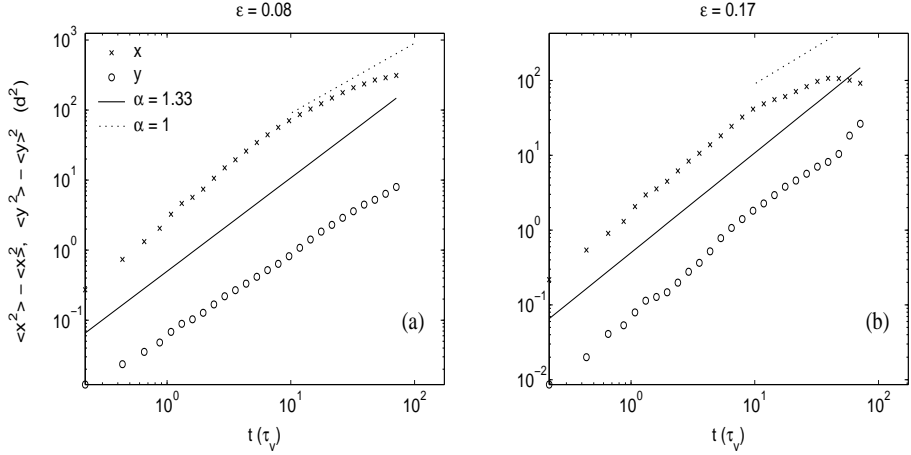, width=12cm, height=6cm}
\caption{Measured anomalous diffusion of defects for two different values
of the non-dimensional temperature differences $\epsilon$ between upper and lower plate.}
\end{figure}
%figDefect turbulence thus may serve as an example where generalized
%versions of statistical mechanics are not only mathematically
%beautiful but physically useful. Note that this is an application
%to a physical system far from equilibrium, where ordinary
%statistical mechanics has little to say.

The simple superstatistical model of section 5 with fluctuating
effective friction $\gamma$ makes sense as a very simple model
for the defect velocity $v$. While ordinary Brownian particles
have constant damping due to Stokes' law $\gamma = \frac{6\pi \nu
\rho a}{m}$, where $\nu$ is the kinematic viscosity of the
liquid, $\rho$ is its density, $m$ is the mass of the particle
and $a$ is the radius of the particle, defects are no ordinary
particles: They are nonlinear excitations and have neither a
well-defined mass $m$ nor a well-defined radius $a$. Thus one
expects that there is an ensemble of damping constants $\gamma$
which depend on the topology of the defect and its fluctuating
environment. In particular, the fastest velocities result from
circumstances in which the defect is moving in a local
environment with only a very small effective local damping $\gamma$
acting. The driving forces $L(t)$ are hardly damped during such a
time interval, and lead to very large velocities for a limited
amount of time, until another region with another $\gamma$ is
reached. The result are $q$-Gaussians, as shown in Fig.6,
and anomalous diffusion. In good approximation this system is
described by $\chi^2$-superstatistics.

%The experimentally observed value $q\approx 1.46...1.50$ for the
%defect statistics means according to eq.~(\ref{qnn}) that there
%are effectively about three independent degrees of freedom that
%contribute to the fluctuating local defect environment. We do not
%know where these three effective degrees of freedom come from,
%but a very simple picture would be that the fluctuating
%environment of the defect is mainly characterized by the states
%of the three convection rolls that merge when forming a defect.

\section{Statistics of cosmic rays}

Our third example is from high energy physics. We will proceed to
extremely high temperatures, where (similar as in defect
turbulence) particles are created and annihilated in pairs.
%Nonextensive statistical mechanics has been shown to work well for
%reproducing experimentally measured cross sections in $e^+e^-$
%collider experiments \cite{bediaga,e+e-}.
We will apply superstatistical techniques to high energy collision
processes on astrophysical scales, leading to the
creation of cosmic ray particles that are ultimately observed on
the earth. The idea to apply superstatistical techniques to the
measured cosmic ray spectrum was first presented in \cite{cosmic},
based on some earlier work in \cite{cos2}.

Experimental data of the measured cosmic ray energy spectrum are shown
in Fig.~8.
\begin{figure}
\epsfig{file=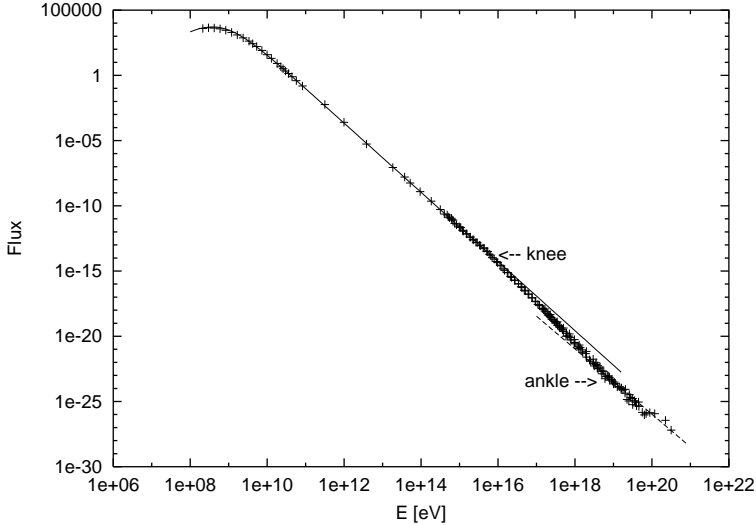} \caption{Observed energy spectrum
of cosmic rays and a fit by eq.~(\ref{can}) with $q=1.215$.}
\end{figure}
Also shown is a curve that corresponds to a prediction of a
superstatistical model. Up to energies of $10^{16}$ eV, the
measured flux rate of cosmic ray particles with a given energy
$E$ is well fitted by a distribution of the form
\begin{equation}
p(E)=C \cdot \frac{E^2}{(1+b(q-1)E)^{1/(q-1)}}. \label{can}
\end{equation}
$E$ is the energy of the particles,
\begin{equation}
E=\sqrt{c^2p_x^2+c^2p_y^2+c^2p_z^2+m^2c^4},
\end{equation}
$b=(k\tilde{T})^{-1}$ is an effective inverse temperature
variable, and $C$ is a constant representing the total flux rate.
For relativistic particles the rest mass $m$ can be neglected and
one has $E\approx c |\vec{p}|$. The distribution (\ref{can}) is a
$q$-generalized relativistic Maxwell-Boltzmann distribution in the
formalism of nonextensive statistical mechanics \cite{tsa1}. The
factor $E^2$ takes into account the available phase space volume.
As seen in Fig.~8, the cosmic ray spectrum is very well fitted by
the distribution (\ref{can}) if the entropic index $q$ is chosen
as $q=1.215$ and if the effective temperature parameter is given
by $k\tilde{T}=b^{-1}=107$ MeV.

The above effective temperature is of the same order of magnitude
as the so-called Hagedorn temperature $T_H$ \cite{hage,e+e-}, an
effective temperature well known from collider experiments. The
Hagedorn temperature is much smaller than the center-of-mass
energy $E_{CMS}$ of a typical collision process and represents a
kind of `boiling temperature' of nuclear matter at the
confinement phase transition. It is a kind of maximum temperature
that can be reached in a collision experiment. Even largest
$E_{CMS}$ cannot produce a larger average temperature than $T_H$
due to the fact that the number of possible particle states grows
exponentially.

Let us now work out the assumption that the power law of the
measured cosmic ray spectrum is due to fluctuations of
temperature. Assume that locally, in the creation process of some
cosmic ray particle, some value of the fluctuating inverse
temperature $\beta$ is given. We then expect the momentum of a
randomly picked particle in this region to be distributed
according to the relativistic Maxwell-Boltzmann distribution
\begin{equation}
p(E|\beta)=\frac{1}{Z(\beta)}E^2 e^{-\beta E}. \label{max}
\end{equation}
Here $p(E|\beta)$ denotes the conditional probability of $E$ given
some value of $\beta$. We neglect the rest mass $m$ so that
$E=c|\vec{p}|$. The normalization constant is given by
\begin{equation}
Z(\beta)=\int_0^\infty E^2 e^{-\beta E} dE=\frac{2}{\beta^3} .
\end{equation}
Now assume that $\beta$ is $\chi^2$-distributed. The observed
cosmic ray distribution at the earth does not contain any
information on the local temperature at which the various
particles were produced. Hence we have to average over all
possible fluctuating temperatures, obtaining the measured energy
spectrum as the marginal distribution
\begin{equation}
p(E)=\int_0^\infty p(E|\beta)f(\beta)d\beta . \label{99}
\end{equation}
The integral (\ref{99}) with $f(\beta)$ given by (\ref{chi22}) and
$p(E|\beta)$ given by (\ref{max}) is easily evaluated and one
obtains eq.~(\ref{can}) with
\begin{equation}
q=1+\frac{2}{n+6} \label{qwert}
\end{equation}
and
\begin{equation}
b=\frac{\beta_0}{4-3q},
\end{equation}
where $\beta_0$ is the average inverse temperature.

The variables $X_i$ in section 3 describe the independent degrees
of freedom contributing to the fluctuating temperature. At very
large center of mass energies, due to the uncertainty relation,
the probed volume $r^3$ is very small, and all
relevant degrees of freedom in this small volume are basically
represented by the 3 spatial dimensions into which heat can flow,
leading to a fluctuating effective temperature in each creation
process of cosmic ray particles.
%We may physically interpret $X_i^2$ as the heat loss in the
%spatial $i$-direction, $i=x,y,z$, during the collision process
%that generates the cosmic ray particle.
%The more heat is lost, the
%smaller is the local $T$, i.e. the larger is the local $\beta$
%given by eq.~(\ref{Gauss}).
The 3 spatial degrees of freedom yield
$n=3$ or, according to eq.~(\ref{qwert}),
\begin{equation}
q=\frac{11}{9}=1.222. \label{qmax}
\end{equation}
For cosmic rays $E_{CMS}$ is very large, hence we expect a
$q$-value that is close to this asymptotic value. The fit in
Fig.~8 in fact uses $q=1.215$, which agrees with the predicted
value in eq.~(\ref{qmax}) to about 3 digits.
%It also coincides
%well with the fitting value $q=1.225$ used by Tsallis et al.
%\cite{cos1} using multi-parameter generalizations of nonextensive
%canonical distributions.

\section{Statistics of train delays}

Our final example leaves the area of classical physics and turns to a more
practical  problem that almost everybody has experienced in the
past. Trains, buses, planes, etc.\ are often delayed! A
statistical analysis of train delay data in the UK was recently
performed in \cite{briggs}. One observes probability densities of
delays that are $q$-exponentials, i.e. that have power-law tails
(see Fig.~9 for an example).

\begin{figure}
\epsfig{file=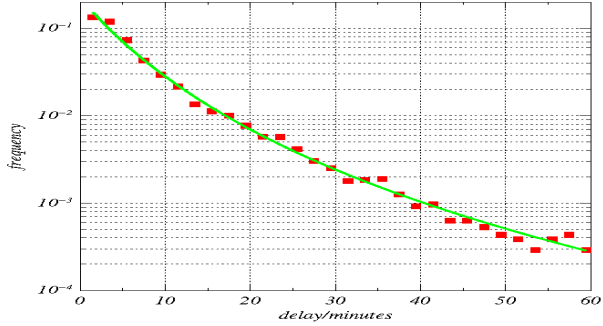, width=10cm, height=8cm} \caption{Delay statistics of
train departures at Swindon station. The trains are heading for London Paddington.}
\end{figure}

We start with a very simple theoretical model for the local departure
statistics of trains. The waiting time distribution until
departure takes place is simply given by that of a Poisson process
\cite{vKa}
\begin{equation}
P(t|\beta)=\beta e^{-\beta t}.
\end{equation}
Here $t$ is the time delay from the scheduled departure time, and
$\beta$ is a positive parameter. The symbol $P(t|\beta)$ denotes
the conditional probability density to observe the delay $t$
provided the parameter $\beta$ has a certain given value.
Clearly, the above probability density is normalized. Large
values of $\beta$ mean that most trains depart very well in time,
whereas small $\beta$ describe a situation where long delays are
rather frequent.

The above simple exponential model becomes superstatistical by
making the parameter $\beta$ a fluctuating random variable as
well. These fluctuations describe large-scale temporal variations
of the British rail network environment. For example, during the
start of the holiday season, when there is many passengers, we
expect that $\beta$ is smaller than usual for a while, resulting
in frequent delays. Similarly, if there is a problem with the
track or if bad weather conditions exist, we also expect smaller
values of $\beta$ on average. The value of $\beta$ is also be
influenced by extreme events such as derailments, industrial
action, terror alerts, etc.

The observed long-term distribution of train delays is then a
mixture of exponential distributions where the parameter $\beta$
fluctuates. If $\beta$ is distributed with probability density
$f(\beta)$, and fluctuates on a large time scale, then one
obtains the marginal distributions of train delays as
\begin{equation}
p(t)=\int_0^\infty f(\beta) p(t|\beta) d\beta = \int_0^\infty
f(\beta) \beta e^{-\beta t}. \label{9}
\end{equation}

Again, a $\chi^2$-distribution of $n$ degrees of freeedom makes sense for $\beta$, leading
to $q$-exponential waiting time distributions of the form
\begin{equation}
%p(t) \sim \frac{1}{\left( 1+b(q-1)t\right)^{\frac{1}{q-1}}}
p(t) \sim {\left( 1+b(q-1)t\right)^{\frac{1}{1-q}}}
\end{equation}
where $q=1+{2}/(n+2)$ and $b= {2}\beta_0 /({2-q})$.
%\begin{equation}
%q=1+\frac{2}{n+2}
%\end{equation}
%and
%\begin{equation}
%b= \frac{2}{2-q} \beta_0.
%\end{equation}
Our model generates $q$-exponential distributions of train delays
by a simple mechanism, namely a $\chi^2$-distributed parameter
$\beta$ of the local Poisson process. This is in good agreement
with the recorded delay data of the British rail network
\cite{briggs}.

% figure made by ./q_beta.sh
%\begin{figure}[ht]
%\begin{center}
%\includegraphics[width=0.8\hsize]{q_beta_plot.ps}
%\caption{The estimated parameters $q$ and $b$ for 23 stations. }
%\label{q_b_plot}
%\end{center}
%\end{figure}

%\setlength{\itemsep}{-1.0ex}\setlength{\parskip}{0mm}
\begin{table}[ht]
\begin{center}
\begin{tabular}{lccl}
\hline\strut
station & $q$ & $b$ & code\\
\hline
Bath Spa& 1.195 & 0.209 & BTH\strut\\
Birmingham & 1.257 & 0.271 & BHM\\
Cambridge & 1.270 & 0.396 & CBG\\
Canterbury East & 1.298 & 0.400 & CBE\\
Canterbury West & 1.267 & 0.402 & CBW\\
City Thameslink & 1.124 & 0.277 & CTK\\
Colchester & 1.222 & 0.272 & COL\\
Coventry & 1.291 & 0.330 & COV\\
Doncaster & 1.289 & 0.332 & DON\\
Edinburgh & 1.228 & 0.401 & EDB\\
Ely & 1.316 & 0.393 & ELY\\
Ipswich & 1.291 & 0.333 & IPS\\
Leeds & 1.247 & 0.273 & LDS\\
Leicester & 1.231 & 0.337 & LEI\\
Manchester Piccadilly & 1.231 & 0.332 & MAN\\
Newcastle & 1.378 & 0.330 & NCL\\
Nottingham & 1.166 & 0.209 & NOT\\
Oxford & 1.046 & 0.141 & OXF\\
Peterborough & 1.232 & 0.201 & PBO\\
Reading & 1.251 & 0.268 & RDG\\
Sheffield & 1.316 & 0.335 & SHF\\
Swindon & 1.226 & 0.253 & SWI\\
York & 1.311 & 0.259 & YRK
\end{tabular}
\caption{The fitted parameters $q$ and $b$ for the departure 
statistics of 23 UK stations.}
\label{q_b_table}
\end{center}
\end{table}

Typical $q$-values obtained from our fits for various stations are in the region
$q=1.15 \dots 1.35$ (see Table~1).
Hence
\begin{equation}
n=\frac{2}{q-1}-2
\end{equation}
is in the region $4\ldots 11$. This means the number of degrees
of freedom influencing the value of $\beta$ is just of the order
we expected it to be: A few large-scale phenomena such as
weather, seasonal effects, passenger fluctuations, signal
failures, repairs of track, etc. seem to be relevant.
In
general, it makes sense to compare stations with the same $q$
(the same number of external degrees of freedom of the network
environment): The larger the value of $b$, the better the
performance of this station under the given environmental
conditions.
%Our analysis shows that two of the best performing
%busy stations according to this criterion are Cambridge and
%Edinburgh.

\setlength{\bibindent}{4mm} % indentation for two digits

%%% Local Variables:
%%% mode: latex
%%% TeX-master: "demofile"
%%% End:
\end{document}